\documentclass[10pt,sigconf,letterpaper]{acmart}
\AtBeginDocument{%
  \providecommand\BibTeX{{%
    \normalfont B\kern-0.5em{\scshape i\kern-0.25em b}\kern-0.8em\TeX}}}

\usepackage{booktabs} 
\usepackage{graphicx}
\usepackage{tikz}
\usepackage{color}
\usepackage{pgfplots}

\usepackage{algorithm}
\usepackage{algorithmicx}
\usepackage{algpseudocode}
\usepackage{cleveref}
\usepgfplotslibrary{dateplot}
\newtheorem{theorem}{Theorem}
\newtheorem{lemma}[theorem]{Lemma}
\crefname{lemma}{lemma}{lemmas}
\setcitestyle{acmnumeric}

\title[Cyclic Arbitrage in DEXes]{Cyclic Arbitrage in Decentralized Exchanges}

\author{Ye Wang}
\email{wangye@ethz.ch}
\affiliation{%
  \institution{ETH Zurich}
  \city{Zurich}
  \country{Switzerland}
}

\author{Yan Chen}
\email{bitmaster@zju.edu.cn}
\affiliation{%
  \institution{Zhejiang University}
  \city{Hangzhou}
  \country{China}
}

\author{Haotian Wu}
\email{wu558536@stu.xjtu.edu.cn}
\affiliation{%
  \institution{Xi'an Jiaotong University}
  \city{Xi'an}
  \country{China}
}

\author{Liyi Zhou}
\email{liyi.zhou@imperial.ac.uk}
\affiliation{%
  \institution{Imperial College London}
  \city{London}
  \country{United Kingdom}
}

\author{Shuiguang Deng}
\email{dengsg@zju.edu.cn}
\affiliation{%
  \institution{Zhejiang University}
  \city{Hangzhou}
  \country{China}
}

\author{Roger Wattenhofer}
\email{wattenhofer@ethz.ch}
\affiliation{%
  \institution{ETH Zurich}
  \city{Zurich}
  \country{Switzerland}
}



\begin{abstract}

Decentralized Exchanges (DEXes) enable users to create markets for exchanging any pair of cryptocurrencies. 
The direct exchange rate of two tokens may not match the cross-exchange rate in the market, and such price discrepancies open up arbitrage possibilities with trading through different cryptocurrencies cyclically.
In this paper, we conduct a systematic investigation on cyclic arbitrages in DEXes. We propose a theoretical framework for studying cyclic arbitrage. With our framework, we analyze the profitability conditions and optimal trading strategies of cyclic transactions.
We further examine exploitable arbitrage opportunities and the market size of cyclic arbitrages with transaction-level data of Uniswap V2. 
We find that traders have executed 292,606 cyclic arbitrages over eleven months and exploited more than 138 million USD in revenue.
However, the revenue of the most profitable unexploited opportunity is persistently higher than 1 ETH (4,000 USD), which indicates that DEX markets may not be efficient enough.
By analyzing how traders implement cyclic arbitrages, we find that traders can utilize smart contracts to issue atomic transactions and the atomic implementations could mitigate users' financial loss in cyclic arbitrage from the price impact.

\end{abstract}

\begin{document}

\maketitle

\section{Introduction}

Decentralized finance (DeFi) in the blockchain ecosystem has attracted a surge of interest with a total gross value locked (GVL) of up to ~$120$ billion USD by October 2021~\cite{lv}. DeFi applications are smart contracts deployed on blockchains, which support a wide variety of financial services~\cite{qin2021cefi,liu2020first}. One of the important applications is decentralized exchange (DEX).
DEXes allow users to exchange their crypto-assets and ensure decentralized and non-custodial trading. Traders sell and buy assets on DEXes by interacting with smart contracts on the blockchain with no centralized authority being involved.
In an automated market maker (AMM) DEX, the exchange rate of each trade is determined by predefined algorithms and market liquidity reserves. Most prominent DEXes in the blockchain ecosystem are AMM DEXes, such as Uniswap~\cite{uni2} and SushiSwap~\cite{sushi}.

\begin{figure}[h]
\centering
\includegraphics[width=0.42\textwidth]{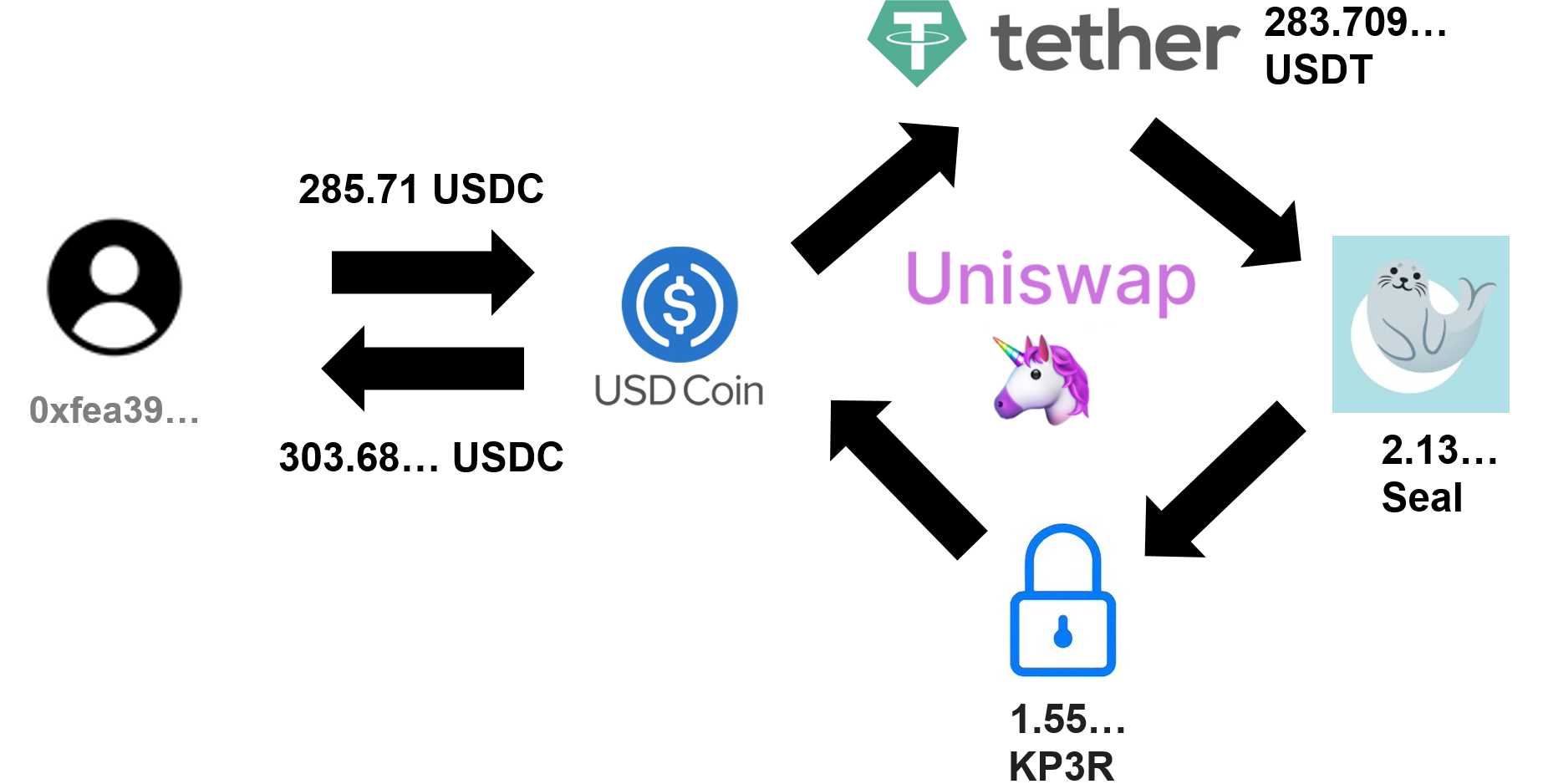}
\caption{An example cyclic arbitrage on Uniswap V2: The trader traded 285.71 USDC through four different exchange markets across USDC, USDT, Seal, Kp3r, and finally received 303.68 USDC. The revenue of this cyclic arbitrage is 17.97 USDC (18 USD).}
    \label{fig:ring}
\end{figure}

The exchange rates between different pairs of cryptocurrencies (also known as tokens) in AMM DEXes are not always in sync perfectly, which opens up arbitrage possibilities for cyclic trading. Assume we have three tokens $A, B, C$, and three markets between any two tokens: $A\rightleftharpoons B, B\rightleftharpoons C$ and $C\rightleftharpoons A$. Traders are able to trade tokens cyclically: Traders can exchange $A$ for $B$ in $A\rightleftharpoons B$, then $B$ for $C$ in $B\rightleftharpoons C$, and finally $C$ for $A$ in $C\rightleftharpoons A$ again, to benefit themselves from the price discrepancies. The combination of these three trades of tokens is a cyclic arbitrage (transaction).\footnote{The particular case with three tokens is called triangular arbitrage~\cite{mccormick1979covered, goldstein1964implications}.} \autoref{fig:ring} shows an example of cyclic arbitrage happened on October 30, 2020.

In this paper, we provide a systematical study on cyclic arbitrage, with (a) theoretical analysis and (b) empirical evaluation with transaction-level data.
We first propose a theoretical framework for studying cyclic arbitrage in AMM. Then, we use the theoretical model to analyze the profitability conditions and optimal trading strategies of cyclic arbitrage in the most common AMM, the constant product market maker (CPMM). We further utilize our theoretical results to investigate market data. We examine exploitable arbitrage opportunities on Uniswap V2~\cite{uni2} from May 4, 2020, to April 15, 2021. 
We find that in almost every block, the revenue of the most exploitable arbitrage opportunity is higher than 1 ETH (4,000 USD). The total revenue of exploitable arbitrage opportunities is persistently higher than 10 ETH over four months, which suggests that the market efficiency of DEXes might be limited.
Later, we measure the market size of exploited cyclic arbitrage in Uniswap V2. We observe 292,606 cyclic transactions over eleven months. Traders receive a total revenue beyond 34,429.11 ETH and pay 8,458 ETH as gas fees. The market size of exploited cyclic arbitrage is significantly smaller than the exploitable opportunities because of the wide variety of tokens and the high gas fees.
Finally, we investigate the implementations of cyclic arbitrage. We find that only 0.03\% of traders submit trades sequentially with multiple blockchain transactions for a single cyclic arbitrage. Alternatively, traders always utilize smart contract technology to submit a cyclic arbitrage within a single blockchain transaction. Such atomic implementations mitigate the financial loss of users in cyclic arbitrage.

To the best of our knowledge, this is the first paper to study the traders' arbitrage behavior of cryptocurrency theoretically and empirically.
We make the following contributions.
First, we provide a systematical understanding of cyclic arbitrage, with the theoretical analysis of the arbitrage model, the measurement of exploitable arbitrage opportunities, the measurement of exploited cyclic arbitrage, and the measurement of cyclic arbitrage implementations.
Second, we provide the measurement of state-of-the-art in terms of (cyclic) arbitrages in DEXes.
Finally, we reveal that blockchain technology enables users to explore more trading strategies. Further studies in DEXes may provide us with novel understandings of user behaviors in financial markets.

\section{Background and Related Work}

\subsection{Ethereum and Smart Contract}

Ethereum is a public blockchain platform, which supports Turing complete functions~\cite{wood2014ethereum}. In contrast to earlier blockchain systems such as Bitcoin, Ethereum allows for decentralized peer-to-peer transactions, and has generic computation capabilities through smart contracts. A smart contract is a set of programs written in high-level languages, e.g., Solidity. Creators compile these programs into executable byte-code and deploy the smart contract on Ethereum.

There are two kinds of accounts in Ethereum: (a) externally owned accounts (EOAs), i.e., those controlled by human users with the corresponding private key, and (b) smart contract accounts (CAs), i.e., those controlled by executable code~\cite{wood2014ethereum}. An Ethereum transaction is broadcast by an EOA to the Ethereum network, while miners then collect transactions and record them in blocks. EOAs can send three types of transactions: \textit{simple transaction}, to transfer the native currency, ETH; \textit{smart contract creation transaction}, to create a new smart contract; and \textit{smart contract execution transaction}, to execute a specific function of a smart contract.

When a transaction is included in blocks by miners, the operation corresponding to the message takes effect. The miner who creates the block modifies the state of corresponding accounts. Each step of the miner's operation consumes a certain amount of gas, and the amount of gas consumed in each block is capped. Users need to specify a gas price for the operation execution when sending transactions. The fee paid by the initiator of transactions to miners is determined by the amount of gas consumed and the gas price (gas fee = gas price $\times$ gas consumption). The miner records receipts of the executed transactions in Ethereum blocks, including the information of the gas fee, the identity of the transaction, and other information generated by smart contracts during the execution.

\subsection{Decentralized Exchanges (DEXes)}

Based on the support of the smart contract, users can create cryptocurrencies on Ethereum in addition to the native coin ETH~\cite{chen2020traveling}. As of October, 2021, more than 450,000 tokens exist on Ethereum platform~\cite{token}. Such development of the blockchain ecosystem incentives the emergence of on-chain financial systems, i.e., Decentralized Finance (DeFi).
DeFi applications are smart contracts deployed on Ethereum, which support sophisticated financial services, such as borrowing and lending~\cite{Compound}, asset exchanges~\cite{uni2}, leverage trading~\cite{alpha-homora-v2}, as well as novel applications such as flash loans~\cite{qin2021attacking}.

Decentralized exchanges (DEXes) are one of the financial infrastructures of the blockchain ecosystem.
Compared to centralized exchanges (CEXes), DEXes do not involve any centralized operators. Users do not need to transfer their assets to the operator before conducting market operations~\cite{harvey2021defi}. In contrast, they send on-chain transactions to DEXes smart contracts to place orders while they keep control of their assets during the entire process.
DEXes support mutual trades among different cryptocurrencies. The prevalent DEX mechanisms operate through so-called automated market makers (AMM), which aggregate liquidity (i.e., \ cryptocurrencies) within liquidity pools contributed by liquidity providers. Traders can exchange cryptocurrencies with the liquidity pool and pay commission fees to the liquidity providers.
For example, when traders want to exchange cryptocurrency $A$ for $B$, they can call a smart contract function that transfers $A$ from the traders' accounts to the liquidity pool between $A$ and $B$. The liquidity pool then sends $B$ to the traders' account. The exchange process does not involve the participation of any other traders. The exchange rate between $A$ and $B$ is determined by transparently predefined functions encoded in the DEX smart contract.

The constant product function is one of the most widely used pricing mechanism. Assume a trader want to exchange $\delta_a$ of $A$ for $B$ token and the liquidity of $A$ and $B$ are $a$ and $b$, respectively. The following equation always hold during the transaction: $a\cdot b = (a+\delta_a\cdot r_1)\cdot(b-\frac{\delta_b}{r_2})$, where $r_1$ and $r_2$ denote the commission fee in asset $A$ and $B$ respectively. In Uniswap~\cite{uni2}, $r_1  = 0.997$ and $r_2=1$, which indicates that the commission fee is equal to $0.3\%\cdot \delta_a$. The remaining liquidity in the pool equals to $(a+\delta_a, b-\frac{r_1\cdot r_2\cdot b\cdot\delta_a}{a+r_1\cdot\delta_a})$.

\subsection{Arbitrage in Cryptocurrency Markets}

Research on arbitrage in cryptocurrencies is still in its beginning.
Previous studies focus on either theoretical analysis of the behavior of miners and traders in blockchain systems~\cite{eyal2015miner, eyal2014majority, kiayias2016blockchain, kwon2017selfish, sapirshtein2016optimal, nayak2016stubborn, koutsoupias2019blockchain, liu2018strategy, marmolejo2019competing, avarikioti2020ride, ramseyerscaling},
or the influence of cryptocurrencies as a potential payment and transaction mechanism in financial markets~\cite{athey2016bitcoin, bohme2015bitcoin, easley2019mining, harvey2016cryptofinance, huberman2017monopoly, pagnotta2018equilibrium}.

Some recent studies~\cite{nan2019bitcoin, pichl2020time, bai2019automated, grimberg2020empirical, makarov2020trading} have noted the prices discrepancies of cryptocurrencies in CEXes. Makarov and Schoar~\cite{makarov2020trading} studied price deviations and potential arbitrage opportunities of Bitcoin, Ethereum, and Ripple for 34 CEXes across 19 countries. They observed significant market segmentation among different countries and suggested that capital controls are the main reasons for market segmentation.
Nan and Kaizoji~\cite{nan2019bitcoin} studied the potential triangular arbitrage with Bitcoin, Euro and U.S. dollar and analyzed market data with a bivariate GARCH model.
Nevertheless, previous studies have only investigated exploitable arbitrage opportunities in CEXes, and none of them have measured exploited arbitrages in the market. Meanwhile, the arbitrage strategy they studied has many constraints to be implemented in markets, such as cross-border capital controls and instantaneous transfers between CEXes.

Recently, DEXes have attracted attention worldwide as an emerging alternative of CEXes for exchanging cryptocurrencies. 
Daian et al.~\cite{daian2020flash,daian2019flash} have analyzed the fundamental weakness of DEXes: slow (on-chain) trading. Since transactions are broadcast in the Ethereum network, adversaries can observe profitable transactions before they are executed and place their own orders with higher fees to front-run the target victim.
Front-running attackers bring threats to the market and system stability. 
Arbitrageurs optimize network latency aggressively and conduct priority gas auctions to front-run profitable trades~\cite{lewis2014flash}, which results in excessive transaction fees affecting normal users in blockchain ecosystems. 
Moreover, because of the high miner-extractable value, fee-based forking attacks and time-bandit attacks are created and bring systemic consensus-layer vulnerabilities. 
Zhou et al.~\cite{zhou2020high} and Torres et al.~\cite{ferreira2021frontrunner} studied sandwiching attacks, i.e., combinations of front and back-running, in DEXes. When observing a victim transaction, attackers place one order just before it (front-run) and place an order just after it (back-run) to benefit themselves through the variance of the exchange rates.
Qin et al.~\cite{qin2021quantifying} quantified the revenue of arbitrages in DEXes. However, this work lacks a systematic analysis of arbitrage behavior and only measures the exploited arbitrage opportunities of 144 cryptocurrencies.

Compared to previous studies, our work fills the following two research gaps to provide a more comprehensive understanding of cyclic arbitrage.
First, we do not only consider potential arbitrage opportunities in the market but also compare with exploited arbitrage opportunities.
Second, we examine different implementations of arbitrage strategies in DEXes and discuss how smart contract technology could help traders to mitigate the financial loss from the price impact.

\section{Cyclic Arbitrage Model}

\label{sec:model}

In this section, we propose a theoretical framework of cyclic arbitrage, and then examine the profitability and the optimal revenue of cyclic arbitrage in CPMM.

\paragraph{Arbitrage model:} A cyclic transaction between $n$ tokens $A_1, A_2, \ldots, A_n$ is a sequence of $n$ trades: 

\begin{itemize}
    \item[\textit{Trade 1}:] Exchange $\delta_1$ of $A_1$ with $\delta_2$ of $A_2$.
    \item[\textit{Trade 2}:] Exchange $\delta_2$ of $A_2$ with $\delta_3$ of $A_3$.
    \item[$\ldots$]
    \item[\textit{Trade $n$}:] Exchange $\delta_n$ of $A_n$ with $\delta_1'$ of $A_1$.
\end{itemize}

Note that the output of \textit{Trade $i$} exactly equals to the input of \textit{Trade $i+1$}, while the revenue of the cyclic transaction is differences between the output of \textit{Trade $n$} and the input of \textit{Trade $1$}, i.e., $\delta_1'-\delta_1$.

As we state in the introduction, the exchange rates between different token pairs may not be synchronized perfectly. However, it is not clear under which conditions that exploitable arbitrage opportunities exist in the market as users also need to pay the commission fee for each trade. Therefore, we analyze the profitability and the optimal trading strategy of a cyclic arbitrage in CPMM.

\paragraph{Profitability conditions:} Assume we have three tokens $A_1, A_2,$ and $A_3$, and three liquidity pools between three tokens. We denote $a_{i,j}$ as the amount of reserved $A_i$ in the liquidity pool with token $A_j$. Then, the revenue of trading $\delta_1$ of token $A_1$ through the cycle $A_1\rightarrow A_2\rightarrow A_3\rightarrow A_1$ is

$
    \delta_1' - \delta_1=(\frac{r_1\cdot r_2\cdot \frac{r_1^2\cdot r_2^2 \cdot a_{2,1}\cdot a_{3,2}\cdot a_{1,3}}{a_{2,3}\cdot a_{3,1}+r_1\cdot r_2\cdot a_{2,1}\cdot a_{3,1}+r_1^2\cdot r_2^2\cdot a_{2,1}\cdot a_{3,2}}}{\frac{a_{1,2}\cdot a_{2,3}\cdot a_{3,1}}{a_{2,3}\cdot a_{3,1}+r_1\cdot r_2\cdot a_{2,1}\cdot a_{3,1}+r_1^2\cdot r_2^2\cdot a_{2,1}\cdot a_{3,2}}+r_1\cdot\delta_1}-1)\cdot \delta_1$.

Traders can benefit from the cyclic transaction only if the revenue, i.e., $\delta_1' - \delta_1$, is larger than 0. 
We generalize the profitability conditions for $n$ tokens in Theorem \ref{lemma:1}. In general words, only if the arbitrage index, i.e., the product of $n$ exchange rates along the cycle, is larger than the commission fees paid in the $n$ pools, there exists exploitable cyclic arbitrage opportunity.

\begin{theorem}[cf.\ Appendix~\ref{sec:ex}]
For a given cycle $A_1\rightarrow A_2\rightarrow \ldots \rightarrow A_n\rightarrow A_{1}$ with $n$ tokens, there exists an arbitrage opportunity for the cyclic transaction if the product of exchange rates $\frac{a_{2,1}\cdot a_{3,2}\cdot \ldots\cdot a_{1,n}}{a_{1,2}\cdot a_{2,3}\cdot \ldots\cdot a_{n,1}}>\frac{1}{r_1^{n}\cdot r_2^{n}}$, where $a_{i,j}$ denotes the liquidity of token $A_i$ in the liquidity pool with token $A_{j}$. Meanwhile, the arbitrage cannot benefit from the reversed direction $A_1\rightarrow A_{n}\rightarrow \ldots \rightarrow A_2\rightarrow A_1$ for cyclic transactions.
\label{lemma:1}
\end{theorem}

\paragraph{Optimal trading strategy:} In addition to noticing whether there are exploitable cyclic arbitrage opportunities in markets with Theorem \ref{lemma:1}, it is also important for traders to find a proper trading strategy to maximize the revenue.
It is intuitive that if the derivative of the revenue function to the initial trading amount is zero, then we acquire the optimal revenue.
We design an algorithm to compute the optimal trading volume of a cyclic arbitrage (cf. Algorithm \ref{alg:1}).
If we exchange $\delta_1$ of $A_1$ through the cycle to obtain $\delta_n$ of $A_n$, we can equate this behavior as exchanging $\delta_1$ of $A_1$ in another liquidity pool between $A_1$ and $A_n$ where the amount of reserved tokens of $A_1$ and $A_n$ are $a_{1,n}'$ and $a_{n,1}'$ respectively. We then further compute the optimal trading volume for the cyclic transaction through $A_1\rightarrow A_2\rightarrow \ldots \rightarrow A_n\rightarrow A_{1}$ as $\delta_a^{op}=\frac{\sqrt{r_1\cdot r_2\cdot a'\cdot a} - a}{r_1}$, where $a=\frac{a_{1,n}'\cdot a_{n,1}}{a_{n,1}+r_1\cdot r_2\cdot a_{n,1}'}$ and $a'=\frac{r_1\cdot r_2\cdot a_{1,n}\cdot a_{n,1}'}{a_{n,1}+r_1\cdot r_2\cdot a_{n,1}'}$. We refer readers to Appendix~\ref{sec:op} for a detailed analysis.

	\begin{algorithm}[tb]
		\caption{Compute the equivalent liquidity of the cycle}
		\begin{algorithmic}[1]
			\State $a_{1,n}'\gets a_{1,2}$
			\State $a_{n,1}'\gets a_{2,1}$
			\For{$i$ from $2$ to $n-1$}
			\State $a_{1,n}'\gets \frac{a_{1,n}'\cdot a_{i,i+1}}{a_{i,i+1}+r_1\cdot r_2\cdot a_{n,1}'}$
			\State $a_{n,1}'\gets \frac{r_1\cdot r_2\cdot a_{n,1}'\cdot a_{i+1,i}}{a_{i,i+1}+r_1\cdot r_2\cdot a_{n,1}'}$
			\EndFor
		\end{algorithmic}
		\label{alg:1}
	\end{algorithm}

\section{Cyclic Arbitrage Opportunities}
\label{sec:oppo}
From Algorithm \ref{alg:1}, we can infer that only the amounts of reserved tokens in liquidity pools and the trading direction are needed for determining the optimal trading strategies.
When an Ethereum block is published, the updated information of liquidity pools is available to all traders.
Therefore, with our theoretical model, we are able to investigate exploitable cyclic arbitrage opportunities in DEXes. We choose Uniswap V2 as the example for the empirical analysis because Uniswap V2 has the longest operation time, the most active traders, and the most liquidity pools among all DEXes~\cite{unipair,rank}.

\paragraph{Data collection:} Every market operation on Uniswap is recorded on Ethereum blocks. We launch go-ethereum, an Ethereum client, on our server to collect data from block 10000835 (when Uniswap V2 has been deployed, 4th May 2020) to block 12244144 (15th April 2021). 
We observe that a \textit{Sync} event is recorded in the blockchain receipt when a market operation happens on Uniswap, and the event includes the liquidity of tokens in the pool after the market operation. Therefore, by collecting all \textit{Sync} events, we are able to recover the market states of all liquidity pools on Uniswap V2 over time. More preciously, we build a dataset that includes the liquidity reserved in all liquidity pools on Uniswap V2 after the execution of transactions in each block.

\paragraph{Profitable opportunities:} With such a dataset, we find profitable cycles in the market and compute the optimal trading input of each exploitable cyclic transaction and the corresponding revenue.
We consider exploitable cyclic arbitrage opportunities within the cycles with length three that involve ETH. As more than 80\% liquidity pools in Uniswap V2 are between ETH and another cryptocurrency~\cite{wang2021behavior}, we infer that we provide a reasonable lower bound for estimating the exploitable opportunities.

\begin{figure}
    \centering
    \includegraphics[width=0.45\textwidth]{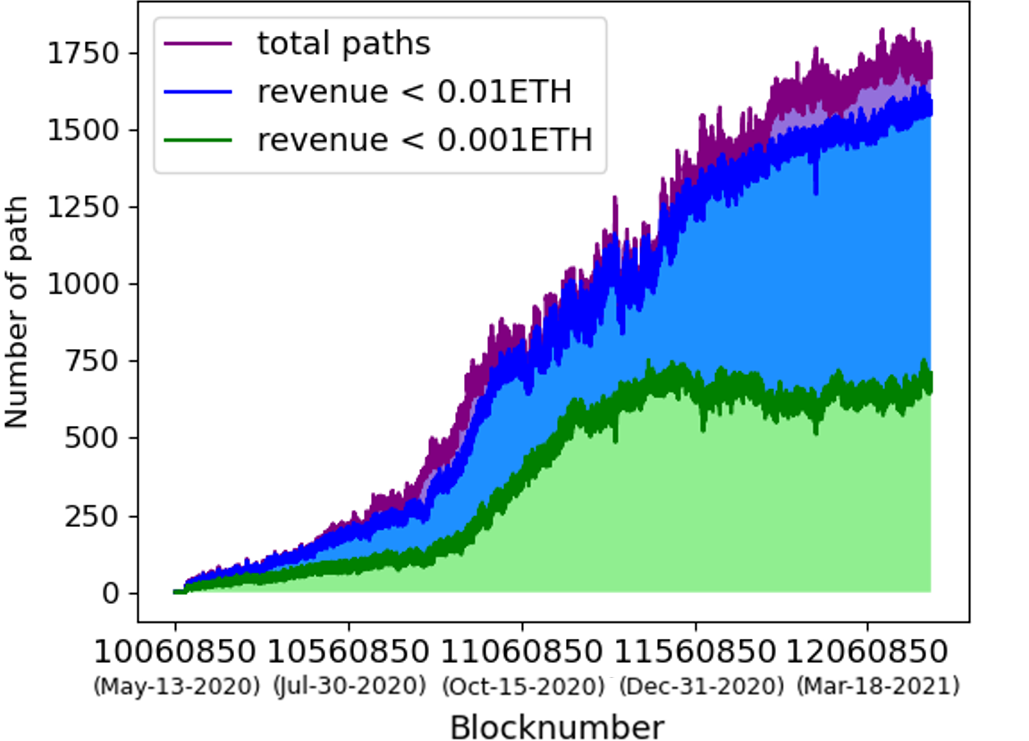}
    \caption{Number of exploitable opportunities in Uniswap V2 over time. The purple line represents the number of cycles that provide revenue higher than 0.0001 ETH. The green represents the number of cycles whose revenue is under 0.001 ETH. The blue line represents the number of cycles whose revenue is under 0.01 ETH.}
    \label{fig:numoppo}
\end{figure}

Because most Ethereum transactions pay more than 0.0001 ETH as gas fee to miners, we count the number of cycles that can provide revenue higher than 0.0001 ETH in \autoref{fig:numoppo}.\footnote{The basic gas consumption of a Uniswap transaction is higher than 100,000 and the gas price is always higher than 1 GWei ($10^{-9}$ ETH per gas).}
The number of exploitable arbitrage opportunities has increased to 1,750 in eleven months. However, we should not overestimate the potential revenue as some cyclic strategies are mutually exclusive. For instance, if two cycles involve the same pair of tokens, only one of them can be exploited with the optimal trading volume.

\begin{figure}
    \centering
    \includegraphics[width=0.45\textwidth]{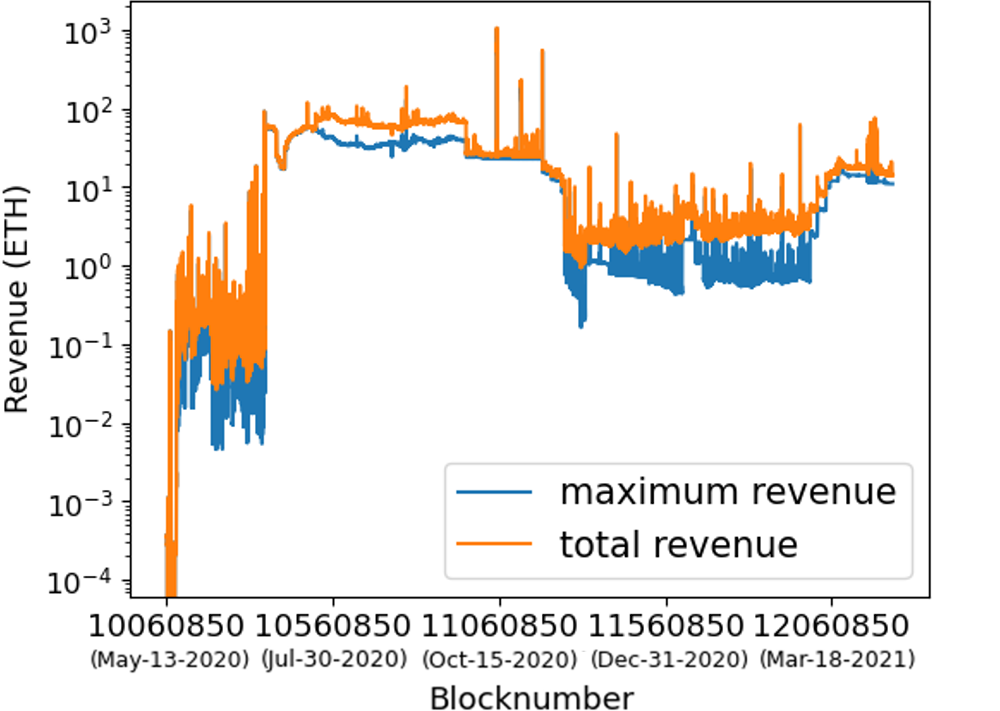}
    \caption{Exploitable revenue of cyclic arbitrage in each block. The blue line represents the maximum revenue of a single arbitrage opportunity, while the orange line represents the sum of a set of independent cyclic arbitrage opportunities.}
    \label{fig:rev}
\end{figure}

Therefore, we consider a set of independent cycles (none of them involves the same liquidity pool) at each block and sum up the revenue of exploiting them, which provides a lower bound of the total revenue that traders can take from the market over time (cf. Figure \ref{fig:rev}). We find that since July 2020, when the market has been developing, the potential revenue of cyclic arbitrage in each block reaches 100 ETH, while the maximum revenue of a single cycle is higher than 10 ETH. However, between November 2020 and March 2021, the total potential revenue per block drops to less than 10 ETH.

\paragraph{Remarks:} Previous studies on arbitrage opportunities of cryptocurrency have mainly focused on CEXes. Compared with their results, we find the following two characteristics of cyclic arbitrage opportunities in DEX.

\textbf{Wider range of arbitrage:} Since CEXes operators will audit the cryptocurrencies trading on the platform, and DEXes support any trading pairs of tokens, the range of the market is wider in DEXes, which further introduces arbitrage opportunities across a broader range of cryptocurrencies. For instance, at the time of writing the paper, Uniswap V2 enables trades between 30,000 tokens while Binance, one of the biggest CEX, only supports less than 400 tokens. We find that more than 2890 liquidity pools and 1143 tokens have been involved in arbitrage opportunities larger than 0.1 ETH (400 USD), which indicates that the range of tokens in cyclic arbitrage is wider in DEXes than CEXes (1,143 vs. 400).

\textbf{Larger market size:} Compared to the arbitrage market in CEXes, the market size in DEXes are larger. As reported by Makarov and Schoar~\cite{makarov2020trading}, the potential arbitrage revenue between 34 CEXes is 2 billion USD over four months (16 million USD per day). However, in a single DEX, i.e., Uniswap V2, as the exploitable revenue of each block exceeds 1 ETH and there are 6,000 blocks per day, the daily revenue is much higher than 24 million USD, even reaching 240 million USD.
Moreover, the arbitrage index in CEXes is always lower than 1.1~\cite{makarov2020trading}. However, the maximum revenue of a single arbitrage opportunity in Uniswap V2 is persistently larger than 1 ETH since July 2020, which suggests that it might be more efficient to exploit arbitrage opportunities in DEXes than in CEXes.

\section{Arbitrage Markets}

After noticing the exploitable arbitrage opportunities in Uniswap V2, we further investigate how many cyclic arbitrages have been executed in the market and how much revenue traders have exploited.

\paragraph{Data collection:} For each successful exchange between two tokens, a \textit{Swap} event will be recorded in the blockchain receipt, and the event includes all trading information, including the input amount of tokens and the output amount tokens. Therefore, we can use the heuristics in Section \ref{sec:model} to determine cyclic transactions. Note that apart from the heuristics mentioned before, we recognize a cyclic transaction such that all trades are executed within 100 blocks (~25 minutes). Because of the volatility in the market price, it is hard to argue that two trades with a time gap of 25 minutes is for exploiting a cyclic arbitrage opportunity.

\begin{figure}
\centering
\setcaptionwidth{0.9\linewidth}
    \begin{tikzpicture}[scale  = 0.8]
\begin{axis}[clip mode=individual, date coordinates in=x,
    xticklabel=\year-\month-\day, xlabel={Date},     ylabel={Number of Transactions},
    log basis y={10}, ymin=0, ymax=4000, 
x label style={
    at={(0.5,-.07)},
    anchor=south,
}, y label style={
    at={(.03,0.1)},
    anchor=west,
}, title style={at={(0.5,1.06)},anchor=north,}
]
\addplot[mark=none, black, line width=1pt] coordinates {
( 2020-05-20,16)
( 2020-05-21,4)
( 2020-05-22,6)
( 2020-05-23,14)
( 2020-05-24,6)
( 2020-05-25,8)
( 2020-05-26,8)
( 2020-05-27,7)
( 2020-05-28,8)
( 2020-05-29,14)
( 2020-05-30,14)
( 2020-05-31,18)
( 2020-06-01,23)
( 2020-06-02,27)
( 2020-06-03,37)
( 2020-06-04,16)
( 2020-06-05,22)
( 2020-06-06,27)
( 2020-06-07,24)
( 2020-06-08,12)
( 2020-06-09,22)
( 2020-06-10,15)
( 2020-06-11,8)
( 2020-06-12,4)
( 2020-06-13,9)
( 2020-06-14,13)
( 2020-06-15,13)
( 2020-06-16,10)
( 2020-06-17,44)
( 2020-06-18,27)
( 2020-06-19,21)
( 2020-06-20,25)
( 2020-06-21,67)
( 2020-06-22,45)
( 2020-06-23,51)
( 2020-06-24,79)
( 2020-06-25,81)
( 2020-06-26,116)
( 2020-06-27,97)
( 2020-06-28,66)
( 2020-06-29,48)
( 2020-06-30,48)
( 2020-07-01,62)
( 2020-07-02,108)
( 2020-07-03,58)
( 2020-07-04,39)
( 2020-07-05,91)
( 2020-07-06,119)
( 2020-07-07,81)
( 2020-07-08,123)
( 2020-07-09,81)
( 2020-07-10,110)
( 2020-07-11,57)
( 2020-07-12,89)
( 2020-07-13,65)
( 2020-07-14,62)
( 2020-07-15,44)
( 2020-07-16,50)
( 2020-07-17,126)
( 2020-07-18,113)
( 2020-07-19,276)
( 2020-07-20,233)
( 2020-07-21,202)
( 2020-07-22,201)
( 2020-07-23,140)
( 2020-07-24,227)
( 2020-07-25,241)
( 2020-07-26,253)
( 2020-07-27,246)
( 2020-07-28,204)
( 2020-07-29,326)
( 2020-07-30,258)
( 2020-07-31,374)
( 2020-08-01,397)
( 2020-08-02,217)
( 2020-08-03,280)
( 2020-08-04,457)
( 2020-08-05,408)
( 2020-08-06,431)
( 2020-08-07,736)
( 2020-08-08,628)
( 2020-08-09,839)
( 2020-08-10,680)
( 2020-08-11,648)
( 2020-08-12,898)
( 2020-08-13,1096)
( 2020-08-14,688)
( 2020-08-15,802)
( 2020-08-16,1158)
( 2020-08-17,964)
( 2020-08-18,1247)
( 2020-08-19,926)
( 2020-08-20,1254)
( 2020-08-21,517)
( 2020-08-22,412)
( 2020-08-23,518)
( 2020-08-24,341)
( 2020-08-25,523)
( 2020-08-26,469)
( 2020-08-27,580)
( 2020-08-28,478)
( 2020-08-29,496)
( 2020-08-30,674)
( 2020-08-31,468)
( 2020-09-01,1206)
( 2020-09-02,799)
( 2020-09-03,880)
( 2020-09-04,876)
( 2020-09-05,709)
( 2020-09-06,787)
( 2020-09-07,899)
( 2020-09-08,914)
( 2020-09-09,565)
( 2020-09-10,718)
( 2020-09-11,1174)
( 2020-09-12,1621)
( 2020-09-13,1292)
( 2020-09-14,1227)
( 2020-09-15,1723)
( 2020-09-16,1459)
( 2020-09-17,572)
( 2020-09-18,1076)
( 2020-09-19,1714)
( 2020-09-20,1907)
( 2020-09-21,1703)
( 2020-09-22,2453)
( 2020-09-23,2101)
( 2020-09-24,1984)
( 2020-09-25,2187)
( 2020-09-26,2009)
( 2020-09-27,1877)
( 2020-09-28,1455)
( 2020-09-29,1170)
( 2020-09-30,1523)
( 2020-10-01,1397)
( 2020-10-02,1334)
( 2020-10-03,1839)
( 2020-10-04,3601)
( 2020-10-05,3630)
( 2020-10-06,3151)
( 2020-10-07,2436)
( 2020-10-08,1827)
( 2020-10-09,1592)
( 2020-10-10,2484)
( 2020-10-11,3025)
( 2020-10-12,1895)
( 2020-10-13,2046)
( 2020-10-14,2189)
( 2020-10-15,1907)
( 2020-10-16,1887)
( 2020-10-17,2314)
( 2020-10-18,2140)
( 2020-10-19,2172)
( 2020-10-20,1601)
( 2020-10-21,1217)
( 2020-10-22,1690)
( 2020-10-23,1855)
( 2020-10-24,1872)
( 2020-10-25,1691)
( 2020-10-26,1920)
( 2020-10-27,1820)
( 2020-10-28,1673)
( 2020-10-29,1493)
( 2020-10-30,1582)
( 2020-10-31,1497)
( 2020-11-01,1425)
( 2020-11-02,1429)
( 2020-11-03,1783)
( 2020-11-04,1950)
( 2020-11-05,1747)
( 2020-11-06,1374)
( 2020-11-07,1582)
( 2020-11-08,1559)
( 2020-11-09,2172)
( 2020-11-10,1727)
( 2020-11-11,1548)
( 2020-11-12,1702)
( 2020-11-13,1510)
( 2020-11-14,1473)
( 2020-11-15,1959)
( 2020-11-16,1503)
( 2020-11-17,1236)
( 2020-11-18,949)
( 2020-11-19,1297)
( 2020-11-20,1088)
( 2020-11-21,1187)
( 2020-11-22,1282)
( 2020-11-23,1246)
( 2020-11-24,1130)
( 2020-11-25,1442)
( 2020-11-26,1343)
( 2020-11-27,1277)
( 2020-11-28,1746)
( 2020-11-29,1721)
( 2020-11-30,1054)
( 2020-12-01,854)
( 2020-12-02,1323)
( 2020-12-03,1544)
( 2020-12-04,1446)
( 2020-12-05,1676)
( 2020-12-06,1945)
( 2020-12-07,1835)
( 2020-12-08,1929)
( 2020-12-09,1475)
( 2020-12-10,1485)
( 2020-12-11,1199)
( 2020-12-12,1293)
( 2020-12-13,1573)
( 2020-12-14,1534)
( 2020-12-15,993)
( 2020-12-16,789)
( 2020-12-17,827)
( 2020-12-18,978)
( 2020-12-19,1342)
( 2020-12-20,1783)
( 2020-12-21,2426)
( 2020-12-22,1566)
( 2020-12-23,1008)
( 2020-12-24,1011)
( 2020-12-25,982)
( 2020-12-26,817)
( 2020-12-27,1159)
( 2020-12-28,754)
( 2020-12-29,701)
( 2020-12-30,673)
( 2020-12-31,651)
( 2021-01-01,754)
( 2021-01-02,797)
( 2021-01-03,701)
( 2021-01-04,742)
( 2021-01-05,675)
( 2021-01-06,588)
( 2021-01-07,505)
( 2021-01-08,580)
( 2021-01-09,655)
( 2021-01-10,797)
( 2021-01-11,840)
( 2021-01-12,856)
( 2021-01-13,820)
( 2021-01-14,734)
( 2021-01-15,825)
( 2021-01-16,939)
( 2021-01-17,943)
( 2021-01-18,934)
( 2021-01-19,768)
( 2021-01-20,803)
( 2021-01-21,804)
( 2021-01-22,866)
( 2021-01-23,835)
( 2021-01-24,942)
( 2021-01-25,562)
( 2021-01-26,726)
( 2021-01-27,750)
( 2021-01-28,842)
( 2021-01-29,666)
( 2021-01-30,469)
( 2021-01-31,827)
( 2021-02-01,596)
( 2021-02-02,573)
( 2021-02-03,937)
( 2021-02-04,763)
( 2021-02-05,704)
( 2021-02-06,755)
( 2021-02-07,829)
( 2021-02-08,792)
( 2021-02-09,623)
( 2021-02-10,573)
( 2021-02-11,843)
( 2021-02-12,788)
( 2021-02-13,689)
( 2021-02-14,742)
( 2021-02-15,756)
( 2021-02-16,709)
( 2021-02-17,875)
( 2021-02-18,923)
( 2021-02-19,722)
( 2021-02-20,652)
( 2021-02-21,618)
( 2021-02-22,584)
( 2021-02-23,571)
( 2021-02-24,624)
( 2021-02-25,543)
( 2021-02-26,628)
( 2021-02-27,618)
( 2021-02-28,642)
( 2021-03-01,577)
( 2021-03-02,570)
( 2021-03-03,722)
( 2021-03-04,761)
( 2021-03-05,1077)
( 2021-03-06,796)
( 2021-03-07,677)
( 2021-03-08,806)
( 2021-03-09,892)
( 2021-03-10,865)
( 2021-03-11,858)
( 2021-03-12,608)
( 2021-03-13,710)
( 2021-03-14,766)
( 2021-03-15,657)
( 2021-03-16,897)
( 2021-03-17,739)
( 2021-03-18,677)
( 2021-03-19,903)
( 2021-03-20,787)
( 2021-03-21,967)
( 2021-03-22,1210)
( 2021-03-23,889)
( 2021-03-24,929)
( 2021-03-25,664)
( 2021-03-26,811)
( 2021-03-27,647)
( 2021-03-28,701)
( 2021-03-29,867)
( 2021-03-30,815)
( 2021-03-31,678)
( 2021-04-01,648)
( 2021-04-02,665)
( 2021-04-03,636)
( 2021-04-04,623)
( 2021-04-05,793)
( 2021-04-06,798)
( 2021-04-07,892)
( 2021-04-08,1433)
( 2021-04-09,953)
( 2021-04-10,810)
( 2021-04-11,777)
( 2021-04-12,709)
( 2021-04-13,689)
( 2021-04-14,617)
( 2021-04-15,257)

  };

\end{axis}

\end{tikzpicture}
    \caption{Daily number of cyclic arbitrages.}
    \label{fig:num}
\end{figure}
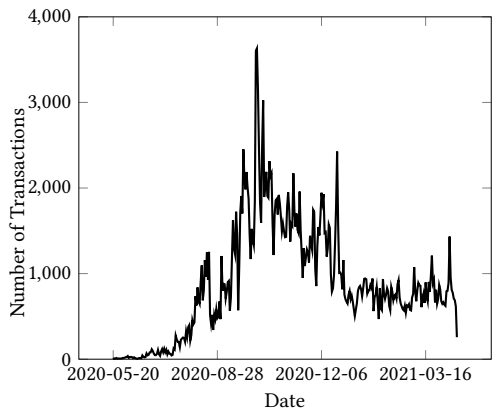

\paragraph{Overall statistics:} Until April 15, 2021, we find 292,606 cyclic transactions, while 287,241 of them start with ETH. These arbitrages happened in 17,189 different cycles. The most popular cycle is ETH-LCX-REVV with 4710 arbitrages. Only 265 cycles have been exploited more than ten times. Around 85\% of cyclic transactions (247,297) are implemented with cycle of length 3.
These observations are consistent with our claims in Section \ref{sec:oppo} that the DEXes have a broad range of cryptocurrencies in arbitrage, and counting the cycle with length three can provide a reasonable lower bound of cyclic arbitrage opportunities.

\paragraph{Transaction number:} \autoref{fig:num} shows the daily number of cyclic transactions in Uniswap. Traders started to perform cyclic arbitrage from May 20, 2020, two weeks after the start of Uniswap V2 (May 04, 2020). 
From May 2020 to September 2020, the market has been growing, reaching more than 3,000 cyclic transactions per day, and then the market enters a relatively active period with more than 1,000 cyclic transactions per day. The market becomes consistent and stable after January 2021, with 600 transactions per day.

\begin{figure}
\centering
    \input{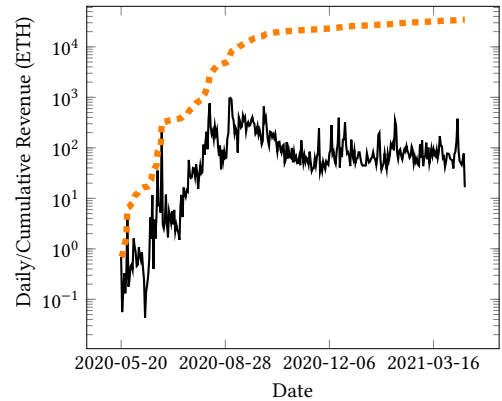}
    \setcaptionwidth{0.9\linewidth}
    \caption{Revenue of cyclic transactions in Uniswap (in ETH). The black line shows a daily market size. The dotted orange line shows the cumulative market size.}
    \label{fig:revenue}
\end{figure}

\paragraph{Transaction revenue:} Because more than 98\% of cyclic transactions start with ETH and there might be error if we sum up the revenue of different cryptocurrencies, we report the revenue of those that start with ETH to provide a reasonable lower bound of the market size. \autoref{fig:revenue} shows the cyclic arbitrage market size, denominated in ETH. The black line shows the daily revenue of cyclic transactions, and the dotted orange line shows the cumulative revenue of cyclic transactions. The total revenue of cyclic arbitrage with ETH in Uniswap V2 is 34,429 ETH.
At the early stage of the market (from May to July 2020), the market of cyclic arbitrage is negligible, i.e., always less than 10 ETH per day.
Later on, the daily arbitrage market has increased to 1000 ETH, while traders become strategic, and the average revenue per transaction increases to 0.2 ETH until September 2020.
Since October 2020, the market has become relatively stable with 100 ETH revenue per day, and the average revenue of each cyclic transaction is around 0.1 ETH. After January 2021, although the daily number of cyclic arbitrage decreases, the total revenue keeps at the same level as before, which indicates that traders can exploit more profitable arbitrage opportunities.

\begin{figure}
    \centering
    \includegraphics[width=0.47\textwidth]{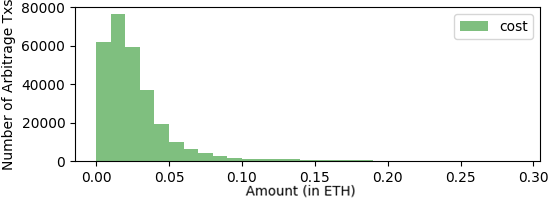}
    \caption{Distribution of gas fee per transaction.}
    \label{fig:cos}
\end{figure}
\begin{figure}
    \centering
    \includegraphics[width=0.35\textwidth]{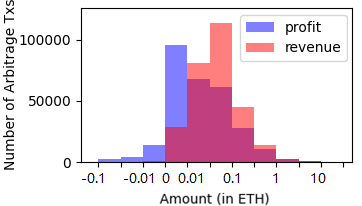}
    \caption{Distribution of profit and revenue per transaction.}
    \label{fig:hist}
\end{figure}

\paragraph{Transaction cost:} Apart from commission fees paid to liquidity providers that have been considered in our theoretical model, traders also need to pay gas fees to miners for executing their market operations.
\autoref{fig:cos} shows the distribution of the gas fee of cyclic transactions.
Around half of transactions (48\%) cost less than 0.02 ETH as the gas fee. The total gas fee (8,458 ETH) accounts for 24.6\% of the total revenue (34,429.11 ETH) of cyclic transactions. \autoref{fig:hist} compares arbitrage income and net profit for cyclic arbitrages. Only 7.1\% of cyclic arbitrages result in negative profits. Gas fees drive revenue per cyclic arbitrage from 0.01-0.3 ETH gross revenue to 0-0.1 ETH net profit, where the majority of cyclic arbitrages (78.4\%) falling in this range.

\paragraph{Remarks:} Compared to arbitrage opportunities in the market, we find that although exploitable arbitrage opportunities exist in almost every block, there are much fewer cyclic arbitrage transactions happening in the market, which might indicate that DEXes are not efficient enough. Since the DEXes are fully built on the blockchain systems, there is a limitation of trading volume on the market. 
Moreover, because of the large size of DEXes, it will take traders time to find cycles with revenue to cover gas fees, as the gas fee of half cyclic arbitrage exceeds 0.02 ETH. 

\section{Arbitrage Implementations}

After studying exploited cyclic arbitrage opportunities, we continue to analyze how users implement their cyclic arbitrage strategies in the market.
In this section, we study the implementations of cyclic arbitrage in DEXes from two perspectives. We first analyze how traders interact with smart contracts to mitigate the financial loss in cyclic arbitrage. Then, we examine the success rate in exploiting arbitrage opportunities of different methods.

\paragraph{Sequential implementation and atomic implementation:}
In traditional CEXes markets, after observing arbitrage opportunities, traders may directly submit $n$ separate orders to realize the cyclic arbitrage. In DEXes, traders have different implementations with smart contract technology: sequential implementation and atomic implementation. Sequential implementation is similar to arbitrage implementations in CEXes. Traders submit $n$ orders separately with $n$ blockchain transactions, while these $n$ transactions are executed sequentially. On the other hand, traders can deploy smart contracts to group all the $n$ trades of a cyclic arbitrage into one blockchain transaction to realize atomic implementations. These $n$ trades will be executed atomically, and no other transactions can be inserted in the middle of them. Moreover, during the execution of smart contract, if some conditions have not been met, then Ethereum allows the smart contract to cancel the execution and revert the system state.

Our first observation of the arbitrage implementation is that 292,518 cyclic transactions are submitted within a single blockchain transaction and execute atomically, while only 88 cyclic arbitrages are conducted sequentially with different blockchain transactions, which indicates that atomic implementation of cyclic arbitrage dominates the entire market.

Because the order of blockchain transaction execution is determined by miners, there might be some other transactions executing in the middle of arbitrage trades if they are submitted separately. These external transactions might change the market price during the arbitrage process and generate the price impact. Consequently, 46 out of 88 sequential implementations of cyclic arbitrage have a negative revenue.
In the scenario of cyclic arbitrage within a single transaction, if the output of the last trades is smaller than the input of the first trade because of the price impact of other transactions before the arbitrage execution, the execution of the $n$ trades can be canceled, while the market states are reverted to those before the arbitrage. Although traders still pay the gas fee to the miner, they do not lose additional money for the non-profitable trades. Only 0.3\% of atomic implementations have a negative revenue, which has a better performance than sequential implementations (52.3\% of transactions have negative revenue).

\paragraph{Successful rate of atomic implementations:}
If the output of the cyclic transaction is smaller than the input, then the smart contract can cancel the execution of the transaction and result in a failed transaction recorded in the blockchain. Traders also need to pay a gas fee to miners for these failed transactions, which should be considered as the cost of cyclic arbitrage. Therefore, we investigate the success rate of atomic implementations.

For all failed transactions, there is no \textit{Swap} event recorded in the blocks. Therefore, it is challenge to determine whether the transaction is aimed at cyclic arbitrage. We consider two methods to construct an upper bound and a lower bound for the success rate for each trader. For the lower bound, we just simply assume that all failed transactions issued by the traders who have ever placed a cyclic transaction are failed implementations of cyclic arbitrage. For the upper bound, we use a replay algorithm to evaluate all failed transactions~\cite{qin2021quantifying}. As the cyclic transaction might be failed because of the price fluctuations between submitting and executing the cyclic transaction, we replay the failed transaction upon its previous block state. If the result of replaying the failed transaction with a non-influenced market state is a cyclic transaction, then we determine it as a failed implementation of cyclic arbitrage. With the upper and lower bounds, we can estimate the success rate of atomic implementations.

In general, traders conduct atomic implementations with two methods: deploying a private smart contract to call exchange functions of Uniswap (282,563 transactions, 840 smart contracts) and directly calling the Uniswap public smart contract (9,955 transactions). Because each trader may have several different EOAs, while only authenticated EOAs can call private smart contracts, we consider all EOAs who use the same smart contract to perform cyclic arbitrage belong to the same trader.

\begin{figure}
    \centering
    \includegraphics[width=0.47\textwidth]{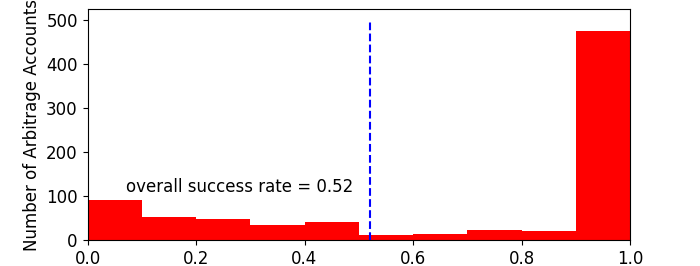}
    \caption{Lower bound of success rate of traders. The dotted line show the overall success rate of all traders.}
    \label{fig:ca}
\end{figure}
\begin{figure}
    \centering
    \includegraphics[width=0.47\textwidth]{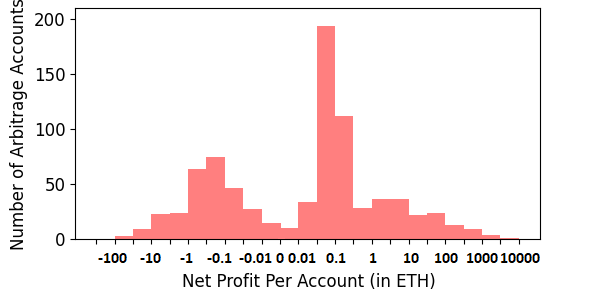}
    \caption{Lower bound of net profit of traders.}
    \label{fig:cap}
\end{figure}

\textbf{Private smart contracts:} We find that traders have called those private cyclic arbitrage smart contract 540,054 times in total and 282,563 successful transactions have been observed. The \textit{lower bound} of the overall success rate is 52\%, while most traders (475 out of 840) have a success rate higher than 90\% (cf. Figure \ref{fig:ca}). If we consider the \textit{upper bound} of success rates, then the overall success rate is 89.6\% as only 34,236 failed transactions issued by arbitrage smart contracts can be recognized as cyclic transaction attempts with our replay experiments. The number of traders who have an upper bound success rate beyond 90\% is 726.
\autoref{fig:cap} shows the distribution of the lower bounds of net profit of 840 smart contracts. The net profit (including failed attempts) of cyclic arbitrageurs is 21,360 ETH (upper bound: 25,742 ETH), while 66\% (upper bound: 77.5\%) of  have a positive balance. Moreover, for those experienced traders who have performed more than 100 cyclic arbitrages, 119 (upper bound: 127) out of 129 traders have a positive net profit, suggesting that cyclic arbitrage is profitable over the long term for experienced traders.

\textbf{Public smart contracts:} Those traders who use public functions to perform cyclic transactions have a much lower success rate than those who deploy private smart contracts. Because the smart contract is public so we can decode their transactions although it is failed executed on the blockchain, which provides us an accurate success rate.
They made 36,492 arbitrage attempts interacting with the Uniswap smart contract, yielding a success rate of 27.3\% (9,955 out of 36,492). Moreover, traders cost 224 ETH as gas fees for failed transactions, which is much higher than their cyclic arbitrage profit (192 ETH) with the public smart contract.

\paragraph{Remarks:}
Atomic implementations of cyclic arbitrage provide two advantages over sequential implementations. First, traders can ensure the completeness of their arbitrage transactions. Either all trades will be executed, or none of them will be executed. Although we only find 88 complete cyclic arbitrage with sequential implementations, there might be more attempts which are failed in the middle of the process and cannot be identified by us. Second, if the cycle is not profitable at the time of the transaction execution, the smart contract can revert to the state when the cyclic trades have not been executed, which mitigates the financial loss when conducting cyclic arbitrage. These two advantages motivate traders to implement arbitrage transactions atomically with smart contracts.

Regarding the failure of cyclic transactions, there can be two reasons: the price impact and front-running attacks~\cite{daian2020flash}. Some traders copy others' arbitrage strategies by observing these transactions in the blockchain P2P network and then create their own transactions in front of the original one to grab the benefits. 
By noticing the significant difference in the success rate, we may conjecture that the resistance to front-running determines the success rate of the private and public smart contracts as the price impact can be considered common to all traders. However, this conjecture needs to be validated more precisely.

\section{Conclusion}

In this paper, we provide a systematic study on cyclic arbitrage in AMM DEXes, including the arbitrage model, arbitrage opportunities, arbitrage markets, and arbitrage implementations.
Additional to understanding cyclic arbitrages in DEXes, this paper also provides insights into two other areas. We compare the arbitrage markets in DEXes and CEXes and discuss the differences between the two exchanges, which may benefit further study in understanding trader behaviors in cryptocurrency markets.
Second, we examine the advantages of smart contract technologies in DEXes markets, which may inspire further studies to explore how traders may utilize blockchain technologies in financial markets.

This paper focuses on the fundamental understandings of cyclic arbitrage in DEXes and raises many interesting and important questions. For example: how do traders decide their strategies; how traders' strategies are affected by average gas fees and over-time variation in congestion; how do cyclic arbitrages influence other market participants; and how do traders combine utilize other services in the blockchain ecosystem to exploit cyclic arbitrage opportunities? We would like to investigate these follow up research questions in future work to promote understanding of user behaviors in the blockchain ecosystem.

\newpage
\bibliographystyle{ACM-Reference-Format}
\bibliography{sample-bibliography}

\appendix
\clearpage
\section{Existence of Cyclic Arbitrage Opportunities}

\label{sec:ex}

\noindent\textbf{\autoref{lemma:1}.} \textit{For a given cycle $A_1\rightarrow A_2\rightarrow \ldots \rightarrow A_n\rightarrow A_{1}$ with $n$ tokens, there exists an arbitrage opportunity for the cyclic transaction if the product of exchange rates $\frac{a_{2,1}\cdot a_{3,2}\cdot \ldots\cdot a_{1,n}}{a_{1,2}\cdot a_{2,3}\cdot \ldots\cdot a_{n,1}}>\frac{1}{r_1^{n}\cdot r_2^{n}}$, where $a_{i,j}$ denotes the liquidity of token $A_i$ in the liquidity pool with token $A_{j}$. Meanwhile, the arbitrage cannot benefit from the reversed direction $A_1\rightarrow A_{n}\rightarrow \ldots \rightarrow A_2\rightarrow A_1$ for cyclic transactions.}
\\
\\
We first give a general expression of cyclic transactions with more than three tokens.

\begin{lemma}
For a cyclic transaction through a path with $n$ edges $A_1\rightarrow A_2\rightarrow \ldots \rightarrow A_n\rightarrow A_{1}$, the first deviation of the utility function at $\delta_1=0$ is $\frac{\partial U}{\partial \delta_1}\bigg|_{\delta_1 = 0}=\frac{r_1^{n}\cdot r_2^{n}\cdot a_{1,n}\cdot a_{2,1}\cdot \ldots\cdot a_{n,n-1}}{a_{1,2}\cdot a_{2,3}\cdot \ldots\cdot a_{n,1}}-1$.
\end{lemma}

\begin{proof}
We can observe that the transaction through $A_1\rightleftharpoons A_2$ and $A_2\rightleftharpoons A_3$ is equivalent to an exchange of $\delta_1$ through liquidity pool $A_1'\rightleftharpoons A_3'$, where $a_{1,3}' = \frac{a_{1,2} \cdot a_{2,3}}{a_{2,3}+r_1\cdot r_2\cdot a_{2,1}}$, and $a_{3,1}'=\frac{r_1\cdot r_2\cdot a_{2,1} \cdot a_{3,2}}{a_{2,3}+r_1\cdot r_2\cdot a_{2,1}}$.

We prove this lemma by induction. We first show it is correct when $n=3$.

We take the utility function of a cyclic transaction of three tokens and compute the first deviation of the function,

\begin{equation}
    \frac{\partial U_{A_1A_2A_3A_1}}{\partial \delta_1}\bigg|_{\delta_1 = 0}=\frac{r_{1}^3\cdot r_{2}^3\cdot a_{1,3}\cdot b_{2,1}\cdot b_{3,2}}{a_{1,2}\cdot a_{2,3}\cdot b_{3,1}} -1
\end{equation}

For $n\geq3$, the inductive hypothesis is that the equation is true for $n$: 

$\frac{\partial U}{\partial \delta_1}\bigg|_{\delta_1 = 0}=\frac{r_1^{n}\cdot r_2^{n}\cdot a_{1,n}\cdot a_{2,1}\cdot \ldots\cdot a_{n,n-1}}{a_{1,2}\cdot a_{2,3}\cdot \ldots\cdot a_{n,1}}-1$.

The inductive step is to prove the equation for $n+1$:

If we exchange $\delta_1$ through $A_1\rightleftharpoons A_2$ pool to get $\delta_2$, and obtain $\delta_3$ from $A_2\rightleftharpoons A_3$ by trading $\delta_2$ right after, these two atomic transactions can is equivalent to a single transaction with $\delta_1$ in a virtual pool $A_1'\rightleftharpoons A_3'$, where $a_{1,3}' = \frac{a_{1,2} \cdot a_{2,3}}{a_{2,3}+r_1\cdot r_2\cdot a_{2,1}}$, and $a_{3,1}'=\frac{r_1\cdot r_2\cdot a_{2,1} \cdot a_{3,2}}{a_{2,3}+r_1\cdot r_2\cdot a_{2,1}}$.

We assume that the statement is correct for any $n$-path cyclic transaction. Consider a cyclic transaction through a $n+1$-path $A_1\rightarrow A^2\rightarrow  \ldots \rightarrow A_n\rightarrow A_{n+1}\rightarrow A_1$, which is equivalent to a $n$-path cyclic transaction through $A_1'\rightarrow A_3'\rightarrow \ldots \rightarrow A_{n+1}\rightarrow A_1'$, where the first deviation of the utility function at $\delta_1=0$ is,

\begin{equation}
    \begin{split}
        \frac{\partial U}{\partial \delta_1}\bigg|_{\delta_1 = 0}&=\frac{r_1^{n}\cdot r_2^{n}\cdot a_{1,n+1}\cdot a_{3,1}'\cdot \ldots\cdot a_{n+1,n}}{a_{1,3}'\cdot a_{3,4}\cdot\ldots\cdot a_{n+1,1}}-1\\
        &=\frac{r_1^{n}\cdot r_2^{n}\cdot a_{1,n+1}\cdot \frac{r_1\cdot r_2\cdot a_{2,1} \cdot a_{3,2}}{a_{2,3}+r_1\cdot r_2\cdot a_{2,1}}\cdot \ldots\cdot a_{n+1,n}}{\frac{a_{1,2} \cdot a_{2,3}}{a_{2,3}+r_1\cdot r_2\cdot a_{2,1}}\cdot a_{3,4}\cdot\ldots\cdot a_{n+1,1}}-1\\
        &=\frac{r_1^{n+1}\cdot r_2^{n+1}\cdot a_{1,n+1}\cdot a_{2,1}\cdot \ldots\cdot a_{n+1,n}}{a_{1,2}\cdot a_{2,3}\cdot \ldots\cdot a_{n+1,1}}-1
    \end{split}
\end{equation}

By the principle of mathematical induction: the first deviation of the utility function at $\delta_a=0$ is $\frac{r_1^{n+1}\cdot r_2^{n+1}\cdot a_{1,n+1}\cdot a_{2,1}\cdot \ldots\cdot a_{n+1,n}}{a_{1,2}\cdot a_{2,3}\cdot \ldots\cdot a_{n+1,1}}-1$, for any $n\geq 3$.

\end{proof}

\begin{proof}[Proof of \autoref{lemma:1}]

Assume that we would like to exchange $\delta_1$ of token $A_1$ through the cycle. When $\delta_1 = 0$, both the value of $U_{A_1A_2A_nA_1}$ and the value of $U_{A_1A^nA^2A_1}$ are 0.

Then we consider the first and the second derivative of the utility function at $\delta_1 = 0$.

\begin{equation}
    \frac{\partial U_{A_1A_2A_nA_1}}{\partial \delta_1}\bigg|_{\delta_1 = 0}=\frac{r_1^{n}\cdot r_2^{n}\cdot a_{1,n}\cdot a_{2,1}\cdot \ldots\cdot a_{n,n-1}}{a_{1,2}\cdot a_{2,3}\cdot \ldots\cdot a_{n,1}}-1
\end{equation}

If arbitrageurs can make profit in this trading direction, then, 

\begin{equation}
    \begin{split}
        0&<\frac{\partial U_{A_1A_2A_nA_1}}{\partial \delta_1}\bigg|_{\delta_1 = 0}\\
        \frac{1}{r_1^{n}\cdot r_2^{n}}&<\frac{a_{1,n}\cdot a_{2,1}\cdot \ldots\cdot a_{n,n-1}}{a_{1,2}\cdot a_{2,3}\cdot \ldots\cdot a_{n,1}}
    \end{split}
\end{equation},

which implies that $\frac{r_1^{n}\cdot r_2^{n}\cdot a_{1,2}\cdot a_{2,3}\cdot \ldots\cdot a_{n,1}}{a_{1,n}\cdot a_{2,1}\cdot \ldots\cdot a_{n,n-1}}<1$, 

then $\frac{\partial U_{A_1A_nA_2A_1}}{\partial \delta_1}\bigg|_{\delta_1 = 0}<0$.

By computing the second derivative of $U_{A_1A_nA_2A_1}$, we know that $\frac{\partial^2 }{\partial \delta_1^2}U_{A_1A_nA_2A_1}$ is negative for all $\delta_1\in R^+$. $U_{A_1A_nA_2A_1}$ is a monotone decreasing function in its domain, and the maximum value of $U_{A_1A_nA_2A_1}$ is 0 at $\delta_1=0$. Therefore, there is no opportunity for arbitrage through the reversed direction $A_1\rightarrow A^n\rightarrow \ldots \rightarrow A^2\rightarrow A_1$ for cyclic transactions.

Arbitrageurs cannot benefit from trading $\delta_1$ through $A_1\rightarrow A^n\rightarrow \ldots \rightarrow A^2\rightarrow A_1$, sequentially. Therefore, arbitrageurs have at most one direction to benefit themselves with cyclic transactions.

\end{proof}

\section{Optimal Trade Volume}

\label{sec:op}
We start from a simple case where cyclic transactions happen in a cycle of token $A_1$, $A_2$, and $A_3$. We denote $a_{i,j}$ as the liquidity of token $A_i$ in the liquidity pool with token $A_j$. 
If we directly exchange $A_3$ with $\delta_1$ of $A_1$, then we receive the output amount of $A_3$, $\delta_3=\frac{r_1\cdot r_2\cdot a_{3,1}\cdot\delta_1}{a_{1,3}+r_1\cdot\delta_1}$.
If we take token $A_2$ as an intermediate, then the output amount of $A_3$ is $\delta_3' = \frac{r_1^2\cdot r_2^2\cdot a_{2,1} \cdot a_{3,2}\cdot\delta_1}{a_{1,2} \cdot a_{2,3}+r_1\cdot\delta_1\cdot(a_{2,3}+r_1\cdot r_2\cdot a_{2,1})}$.

Compare $\delta_3$ and $\delta_3'$, $\delta_3'$ can be written in the format as $\delta_3'=\frac{r_1\cdot r_2\cdot\frac{r_1\cdot r_2\cdot a_{2,1}\cdot a_{3,2}}{a_{2,3}+r_1\cdot r_2 \cdot a_{2,1}}\cdot \delta_1}{\frac{a_{1,2}\cdot a_{2,3}}{a_{2,3}+r_1\cdot r_2 \cdot a_{2,1}}+r_1\cdot \delta_1}$. Therefore, this transaction through two pools can be considered as a transaction through an equivalent pool of $A_1$ and $A_3$ where the liquidity of $A_1$ is $a_{1,3}' = \frac{a_{1,2}\cdot a_{2,3}}{a_{2,3}+r_1\cdot r_2 \cdot a_{2,1}}$ and the liquidity of $A_3$ is $a_{3,1}' = \frac{r_1\cdot r_2\cdot a_{2,1}\cdot a_{3,2}}{a_{2,3}+r_1\cdot r_2 \cdot a_{2,1}}$.

Then, we can rewrite the utility function of a cyclic transaction in $A_1\rightarrow A_2\rightarrow A_3\rightarrow A_1$ with input $\delta_1$ as

\begin{equation}
    \begin{split}
    U_{A_1A_2A_3A_1} = (\frac{r_1\cdot r_2\cdot\frac{r_1\cdot r_2\cdot a_{2,1}\cdot a_{3,2}}{a_{2,3}+r_1\cdot r_2 \cdot a_{2,1}}\cdot}{\frac{a_{1,2}\cdot a_{2,3}}{a_{2,3}+r_1\cdot r_2 \cdot a_{2,1}}+r_1\cdot \delta_1}-1)\cdot\delta_1
    \end{split}
\end{equation}

Let us denote $a=\frac{a_{1,3}'\cdot a_{3,1}}{a_{3,1}+r_1\cdot r_2\cdot a_{3,1}'}$ and $a' = \frac{r_1\cdot r_2\cdot a_{1,3}\cdot a_{3,1}'}{a_{3,1}+r_1\cdot r_2\cdot a_{3,1}'}$

\begin{equation}
    \begin{split}
    \frac{\partial U_{A_1A_2A_3A_1}}{\partial \delta_a} &= \frac{r_1\cdot r_2\cdot a'\cdot(a+r_1\cdot\delta_1)-r_1\cdot(r_1\cdot r_2\cdot a'\cdot\delta_1)}{(a+r_1\cdot\delta_1)^2}- 1\\
    \frac{\partial U_{A_1A_2A_3A_1}}{\partial \delta_a} &= \frac{r_1\cdot r_2\cdot a'\cdot a}{(a+r_1\cdot\delta_1)^2}- 1
    \end{split}
\end{equation}

To maximize the utilize, we have $\frac{\partial U_{A_1A_2A_3A_1}}{\partial \delta_1} = 0$, therefore,

\begin{equation}
    \begin{split}
    \frac{r_1\cdot r_2\cdot a'\cdot a}{(a+r_1\cdot\delta_1)^2}- 1 &= 0\\
    r_1\cdot r_2\cdot a'\cdot a &= (a+r_1\cdot\delta_1)^2\\
    \delta_1 &= \frac{\sqrt{r_1\cdot r_2\cdot a'\cdot a} - a}{r_1}
    \end{split}
\end{equation}

We can find the equivalent pool of transactions for longer paths between any pair of tokens iteratively as in the first step of computing pools equivalent to $A_1\rightleftharpoons A_2$ and $A_2\rightleftharpoons A_3$ and then determine the optimal input of the cyclic transaction.

\end{document}